\documentclass[12pt]{iopart}
\usepackage{iopams,curvesls}
\usepackage{latexsym}   
\jl{6}

\def\R{\mathbb{R}}
\def\O{\mathbf{O}\,}
\def\interior{\dot}               
\def\closure{\bar}                
\def\restrto#1{|_{\lower1pt\hbox{${}_{#1}$}}} 
\def\setof#1{\lbrace#1\rbrace}    
\def\property{\,|\,}              
\def\setdif{-}                    
\def\cross{\times}                
\def\from{\colon}                 
\def\isomorph{\cong}              
\def\immersion{\hookrightarrow}   
\def\orie{\mathrm{or}}            
\def\coorie{\mathrm{coor}}        
\def\norie{\mathrm{n}}            
\def\genip#1#2#3{\xgenip{#1}#2??{#3}}
\def\xgenip#1#2,#3??#4{#1#2,#3#4}
\def\ip#1{\genip\langle{#1}\rangle}
\def\tens{\otimes}                
\def\id{\mathrm{id}}	          
\def\dual{*}                      
\def\det{\mathrm{det\,}}          
\def\Hodge{\ast}	          
\def\Lie{\mathcal L}	          
\newtheorem{Proposition}{Proposition}[section]
\newtheorem{proposition}[Proposition]{Proposition}

\newtheorem{theorem}[Proposition]{Theorem}
\newtheorem{corollary}[Proposition]{Corollary}

\newtheorem{definition}[Proposition]{Definition}
\newtheorem{example}{Example}
\newtheorem{remark}[Proposition]{Remark}

\def\proof{\noindent\textbf{Proof:}\quad}
\def\qed{\ensuremath{\quad\Box\quad}}
\begin{document}
\title{Charge and the topology of spacetime}
\author{Tammo Diemer\dag \, and Mark J Hadley\ddag\footnote[3]{To
whom correspondence should be addressed.}}

\address{\dag\ Mathematisches Institut, Universit\"at
Bonn, Beringstra{\ss}e 3, D-53115 Bonn, Germany, email:
tammo.diemer@math.uni-bonn.de}

\address{\ddag\ Department of Physics, University of Warwick, Coventry
CV4~7AL, UK, email: Mark.Hadley@warwick.ac.uk}

\begin{abstract}
A new class of electrically charged wormholes is described in which
the outer two sphere is not spanned by a compact coorientable
hypersurface. These wormholes can therefore display net electric
charge from the source free Maxwell's equation. This extends the work
of Sorkin on non-space orientable manifolds, to spacetimes which do
not admit a time orientation. The work is motivated by the suggestion
that quantum theory can be explained by modelling elementary particles
as regions of spacetime with non-trivial causal structure. The
simplest example of an electrically charged spacetime carries a
spherical symmetry.
\end{abstract}

\maketitle

\section{Introduction} \label{introduction}
It is known that the source free equations of electromagnetism can
display apparent charge in regions of spacetime with a non-trivial
topology. The simplest examples are Wheeler's wormholes
\cite{misner_wheeler}, where one mouth has positive electric charge
and the other mouth has the same amount of negative electric
charge. These topological structures of spacetime are commonly known
as geons. Sorkin constructed an example in which the entire wormhole
exhibits magnetic charge \cite{sorkin}. This wormhole does not admit a
space orientation. A purely differential topological discussion of the
notion of magnetic and electric charge in spacetime is presented. It
contains a review and extension of the work of Wheeler, Misner and
Sorkin.

It has been suggested by one of the authors that quantum mechanics
could be explained by modelling elementary particles as $4$-geons
(geons with a non-trivial casual structure) \cite{hadley97}. A second
paper relates topology changing interactions between geons with a
reversal of the time orientation, resulting in spacetimes that are not
time orientable \cite{hadley98}. Previously spacetimes that lacked a
time orientation were considered unphysical; now, from this radical
perspective, they are not only physically relevant but very important.

We apply the source free Maxwell equations and the definition of
electric and magnetic charge to spacetimes which lack space or time
orientations. Various wormholes are constructed with each type of
orientability. In these examples net magnetic charge can appear when
space is not orientable and electric charge when spacetime is not time
orientable. Finally a new spacetime is described which lacks a time
orientation, has spherical symmetry and the outward appearance of a
point source of electric charge - an electric monopole. Topological
obstructions prevent the construction of an analogous magnetic
monopole.

The organisation of the article is as follows: The next section is
applicable to forms and multi vector densities on manifolds of any
dimension. It contains the definition of magnetic and electric charge
and the corresponding conservation laws. In
section~\ref{wormholetopology} we define wormhole topologies and
construct physically relevant examples of four dimensional
spacetimes. They carry two forms and bi vector densities, which
display electric and magnetic charge from the source free Maxwell
equations. A more general topological construction is defined in
section~\ref{linebundle} which leads to the construction of a
spacetime possessing an electric monopole with spherical symmetry.
\section{Electromagnetic charge} \label{chargeinem}
We are concerned here with a closed $k$-form $F$ and a divergence free
$(n-k)$-multi vector density $H$ on a smooth $n$-dimensional manifold
$M$. In formulas we deal with $F \in C^\infty(M,\Lambda^kT^*M)$ and $H
\in C^\infty(M,L^{-n} \tens \Lambda^{n-k} TM)$ subject to the
equations $dF=0$ and $div H=0$. (A review of exterior
calculus is given in the appendix).

For $n=4$ and $k=2$ this reduces to the relativistic version of the
first and second Maxwell equations for electromagnetism. In
electromagnetism $F$ is called the \emph{Faraday two form} (electric
field plus magnetic flux) and $H$ is called the \emph{Faraday bi
vector} (dielectric displacement plus magnetic loop tension). $F$ and
$H$ have to be related by the so called \emph{constitutive relation}
of the matter in question. In vacuum this relation is simply given by
a Lorentzian metric $g$. In index notation this relation reads
$g_{ac}g_{bd}H^{cd} = \sqrt{|\det g|} F_{ab}$. This formula is
invariant under rescaling the metric, which means that the
identification only depends upon the conformal class of $g$. (This
point of view goes back to Mie and is taken from \cite{weyl}).

If a Riemannian or Lorentzian metric is given on $M$, some authors
replace the multi vector density $H$ by another $k$-form $H'$ and use
a (local) Hodge star operator to require $\Hodge d \Hodge H' =
0$. However, we will take advantage of the clear geometric distinction
between forms and multi vector densities, leading to a clear
distinction between the notion of magnetic and electric charge. This
distinction can already be seen in the following definitions: (For the
meaning of the following integrals and the notion of a coorientation,
see the appendix)

\begin{definition}[Magnetic charge]\label{magneticcharge}
Given a closed $k$-form $F$ as above we define the \emph{magnetic
charge} (represented by $F$) contained in an (immersed) oriented
$k$-dimensional sphere $S^k \immersion M$ by
$$Q_m:=\int_{S^k,\orie}F.$$
\end{definition}

\begin{definition}[Electric charge]\label{electriccharge}
Similarly, if $H$ denotes a divergence free $(n-k)$-multi vector
density we define the \emph{electric charge} (represented by $H$)
contained in an (immersed) cooriented $k$-dimensional sphere $S^k
\immersion M$ by: $$Q_e:=\int_{S^k,\coorie}H.$$
\end{definition}

The definition of magnetic (or electric) charge can easily be extended
to an integral over any oriented (cooriented) compact $k$-manifold,
but we will restrict the examples to spheres for convenience. As is
well known, these two quantities satisfy conservation laws, which
follow from Stoke's theorem and the Divergence theorem: Combining the
definition of magnetic charge \ref{magneticcharge} with the first
Maxwell equation $dF=0$, and Stoke's theorem (see \ref{stokesthm}),
gives a vanishing result, or conservation law for magnetic charges:

\begin{corollary}\label{vanishingmagneticcharge}
If $S^k$ is the boundary of an (immersed) compact oriented
$(k+1)$-dimensional submanifold $\Sigma$ with boundary
$\partial\Sigma=S^k$, then $Q_m = 0$.
\end{corollary}

Consequently, if $S^{k},\orie$ and ${S^{k}}',\orie'$ are two immersed
spheres, such that there is an (immersed) compact oriented
$(k+1)$-dimensional submanifold $\Sigma$ with these two spheres as
boundary: $\partial\Sigma,\orie=S^{k},\orie\, \cup\,
{S^{k}}',-\orie'$, then $S^{k}$ and ${S^{k}}'$ contain the same amount
of magnetic charge.

Analogously combining the definition of electric charge
\ref{electriccharge} with the second Maxwell equation $div H=0$, the
Divergence theorem (see \ref{divergencethm}), gives a vanishing
result, or conservation law for electric charges:

\begin{corollary}\label{vanishingelectriccharge}
If $S^k,\coorie$ is the boundary of an (immersed) compact cooriented
$(k+1)$-dimensional submanifold $\Sigma$ with boundary
$\partial\Sigma=S^k$, then $Q_e = 0 $.
\end{corollary}

Consequently, if $S^{k},\coorie$ and ${S^{k}}',\coorie'$ are two
immersed spheres, such that there is an (immersed) compact cooriented
$(k+1)$-dimensional submanifold $\Sigma,\coorie$ with these two
spheres as boundary: $\partial\Sigma,\coorie=S^{k},\coorie\, \cup\,
{S^{k}}',-\coorie'$, then $S^{k}$ and ${S^{k}}'$ contain the same
amount of electric charge.

It is highly significant that while the vanishing of magnetic charge
requires $\Sigma$ to be orientable, the vanishing of electric charge
requires it to be coorientable. A simple $(1+1)$-dimensional example
of a submanifold that is orientable but not coorientable is a circle
going round a M\"obius strip. The circle $S^1$ is orientable, but on
the M\"obius strip a consistent normal vector cannot be defined along
$S^1$. More generally: a surface $\Sigma\immersion M$ of dimension
$n-1$ is coorientable in $M$ if there is a vector field along $\Sigma$
which is everywhere transversal to $T\Sigma$ (hence the boundary of a
manifold is always coorientable). If the manifold $M$ itself is
orientable, then a $k$-dimensional $\Sigma$ is coorientable if and
only if $\Sigma$ is orientable. Similarly, if a Lorentzian manifold
$M^{n-1,1}$ is time orientable and if $\Sigma$ is a spacelike surface
of dimension $n-1$ then $\Sigma$ is always coorientable.

Note, that the above definitions and claims have a pure differential
topological nature. It is irrelevant whether $\Sigma$ is spacelike, or
timelike in places, as in some of the examples in
\cite{hadley98}. However when the assumptions for the two vanishing
results are not satisfied then examples exist of apparent net charge
arising from the source free equations. Such examples will be
constructed in the following two sections.

\begin{remark}
As Sorkin pointed out \cite{sorkin}, one could twist Maxwell's theory
by the orientation bundle (bundle of pseudoscalars): real numbers
would turn into pseudo scalars (and vice versa), $k$-forms $F$ would
turn into $(n-k)$-multi vector densities $F'$, multi vector densities
$H$ would turn into forms $H'$. Maxwell's equation then would read
$div F'=0$ and $dH'=0$, which shows that it is not clear how to
distinguish between those two theories. The integral of $H'$ over an
oriented $k$-sphere would now define electric charge. Sorkin uses this
freedom of writing Maxwell's equations, so he defines electric and
magnetic charge using a convention that differs from the more common
one presented here.
\end{remark}

A clear correspondence between the definition of charge given above
and the notion of a point charge with an associated worldline is
given by the following two examples: The \emph{Coulomb field} of an
electric point charge $q_e\in\R$ with a straight worldline
$\R\cross\setof{0}$ in Minkowski space $\R\cross\R^3$ becomes singular
along the worldline. On $\R\cross\R^{>0}\cross S^2$ it is given by
$E(t,r,\sigma)=1/(4\pi r^2)(0,1,0)$. The field of observers at rest
with the electric charge is $N(t,r,\sigma):=(1,0,0)$.

\begin{example}[Electric charge on an incomplete manifold]
\label{elechargeincommfd}
On Minkowski space with a removed worldline
$\R\cross\R^{>0}\cross S^2$, the radial field $E$ together with the
observer field $N$ induce a smooth Faraday bi vector $H:= q_e E \wedge
N$. The electric charge contained in the two sphere $S^2$ linking the
removed worldline corresponds to $q_e$.
\end{example}

In contrast a magnetic (monopole) point charge can be modelled by a
pseudo scalar $q_m\in L^{4}\tens\Lambda^4{\R^4}^\dual$ (see appendix)
along a worldline.

\begin{example}[Magnetic charge on an incomplete manifold]
\label{magchargeincommfd}
On Minkowski space with removed worldline $\R\cross\R^{>0}\cross S^2$,
the radial field $E$ and the observer field $N$ induce a smooth
Faraday two form $F:= \ip{q_m, E \wedge N}$. The magnetic charge
contained in the two sphere $S^2$ linking the removed worldline
corresponds to $q_m$.
\end{example}

Since the linking two sphere $S^2$ on the resultant manifold
$\R\cross\R^{>0}\cross S^2$ is not the boundary of a compact
hypersurface, the vanishing theorems from above do not apply to the
two examples. However, such spacetimes are geodesically incomplete.
\section{Wormhole topology}\label{wormholetopology}
In order to model an elementary particle as a region of non-trivial
topology we define a {\em topological geon}. A point particle traces
out a world line in spacetime, but if the particle has some size and
structure we can consider a worldtube such that the particle is always
inside the tube. Some properties of a particle may depend upon the
internal structure but others, such as mass, electric charge and
magnetic charge can be determined, and defined, entirely by
measurements made outside the worldtube. It will be shown that certain
topological structures inside the worldtube together with the
source free Maxwell equations exhibit net charge, exactly as if the
worldtube contained a point charge.

Denote by $B^{k+1}\subset \R^{k+1}$ the unit ball in $\R^{k+1}$ with
boundary $S^k$. (The open Ball will be denoted by
$\interior{B}^{k+1}$). We now restrict our attention to $n$-manifolds
$M$ containing a cylinder diffeomorphic to $\R^{n-k-1}\cross S^{k}$,
for $1\le k\le (n-2)$. The region outside the cylinder is assumed to
be ``topologically trivial'' in the sense that it is diffeomorphic to
$\R^n$ minus a solid cylinder. The region inside the cylinder can be
highly non-trivial, but has to be ``complete'' in the sense that the
$k$-sphere $S^k$ is the boundary of a compact
$(k+1)$-surface. Physically relevant are the cases $k=2$ and $n=4$. In
this case the inner region of non-trivial topology can be considered
as a model of an elementary particle a so called \emph{$4$-geon}
\cite{hadley97}.

\begin{definition}[Topological $n$-geon]\label{topological4geon}
We will call a smooth $n$-di\-men\-sional manifold $M$ a
\emph{topological $n$-geon}, if it contains an open set $T\subset M$
(the \emph{worldtube}) such that $M\setdif T$ is diffeomorphic to
$\R^n$ with an open tube removed: $M\setdif T\isomorph\R^{n-k-1}\cross
(\R^{k+1}\setdif{\interior{B}}^{k+1})$ and if the $k$-sphere $S^{k}$
linking this tube $T$ is the boundary of some compact (immersed)
$(k+1)$-dimensional surface $\Sigma\immersion{\closure {T}}$ going
through the worldtube $T$. Such a $\Sigma$ will be called a
\emph{spanning surface}.
\end{definition}

If $M$ is a topological $n$-geon, then the $k$-sphere $S^{k}$ linking
the world tube $T$ is orientable and coorientable. Hence it can be
used to define the magnetic and electric charge of the $n$-geon.

\begin{definition}[Net charge] \label{netcharge}
If a closed $k$-form or a divergence free
$(n-k)$-multi vector density is given on $M$, the charge inside a
$k$-sphere $S^k$ linking the worldtube $T$ is also called the \emph{net
charge} of the $n$-geon.
\end{definition}

Topological $n$-geons with non-trivial topology can be explored using
spacetimes with wormholes. A two dimensional (space like) wormhole
is simply a surface with a handle attached. Mathematically a wormhole
may be constructed by cutting two regions out of spacetime and joining
the edges according to a given rule. A M\"obius strip is constructed
from a cylinder by cutting and then joining the edges, after a
rotation of $\pi$. The rule applied for joining the excised regions
allows all combinations of space and time orientations to be
constructed. The construction differs fundamentally from examples in
the literature because we allow for a non-trivial identification of
time as well as space coordinates. The resulting structures are {\em
topological n-geons} which carry all combinations of net magnetic and
net electric charge, as will be seen in the rest of this section.

A \emph{wormhole} is constructed out of $\R^n$ by removing two open
solid cylinders and identifying points on the boundaries: Choose a
space time split $\R^n =\R^{n-k-1} \cross \R^{k+1}$ and let $B_1
:=B^{k+1}(-a) \subset \R^{k+1}$ and $B_2 :=B^{k+1}(+a)$ be the unit
balls centered at $-a$ and $a\in\R^{k+1}$ (with $|a|>1$). The
identifications at the cylindrical boundaries of
$\R^{n-k-1}\cross(\R^{k+1}\setdif ({\dot{B_1}}\cup{\dot{B_2}}))$ will
be determined by orthogonal maps $\phi\in\O(\R^{n-k-1})$ (time
identification) and $\psi\in\O(\R^{k+1})$ (space identification): for
all $\vec r\in S^k$ we identify $(t,{\vec r}-a)\sim(\phi(t),\psi({\vec
r})+a)$. This defines a topological manifold. The smooth structure
should be such that a curve which moves into one cylinder comes out of
the other: i.e.~a velocity vector $(\dot t,\dot q)$ at $(t,{\vec r}
-a)$ will be identified with $(\phi(\dot t),\psi({\dot q}-2\ip{{\dot
q},{\vec r}}{\vec r}))$ (mirror image rule).

\begin{remark}\label{rem:nometric}
A Lorentzian metric on a manifold induces a light cone structure for
which it makes sense to define space and time orientability of the
manifold. In our topological discussion we don't have a lightcone
structure, so we define space and time orientation for the particular
class of examples we have in mind. However, the examples have been
constructed to facilitate the addition of a Lorentzian metric, and if
one is added then the two definitions should coincide.
\end{remark}

\begin{definition}[Wormhole]
The resulting quotient manifold $M :=
\R^{n-k-1}\cross(\R^{k+1}\setdif( {\dot{B_1}} \cup {\dot{B_2}})) /
(\phi,\psi)$ will be called a \emph{wormhole of signature
$(n-k-1,k+1)$}. $M$ is called \emph{time orientable} if $\det\phi=+1$
and $M$ is called \emph{space orientable} if $\det\psi=-1$.
\end{definition}

This is clearly an example of an $n$-geon. The sphere $S^k(R)\subset
\R^{k+1}$ centered at the origin with large radius $R>|a|+1$ links
both removed tubes and is the boundary of $\Sigma:=B^{k+1}(R)\setdif
({\dot{B_1}} \cup {\dot{B_2}})/\psi$. The vanishing results
\ref{vanishingmagneticcharge} and \ref{vanishingelectriccharge} can be
applied to give:

\begin{proposition}
If the wormhole $M$ is space orientable, then $M$ does not contain net
magnetic charge, since the spanning $\Sigma$ is orientable. If $M$ is
time orientable, then $M$ does not contain net electric charge, since
the spanning $\Sigma$ is coorientable.
\end{proposition}

Wormholes of physical interest have $n=4$ and $k=2$. Those described
in the literature have been restricted to $\phi(t) = t$ and are
therefore time orientable, hence they do not carry electric net charge
(in the sense of definition \ref{electriccharge}). All possible
orientability characteristics are treated in the following examples.
Most cases are well known, they are included for completeness. The
examples are labelled by the static multipole they naturally carry.

\begin{example}[Dipole]\label{dipol}
Misner and Wheeler~\cite{misner_wheeler} used a reflection for the
space identification: $\psi({\vec r}):={\vec r} -2{\ip{{\vec
r},a}\over \ip{a,a}}a$ and $\phi(t)=t$. This wormhole naturally
carries static dipole fields, i.e.~a Faraday bi vector induced by two
opposite electric charges and a Faraday two form induced by two
opposite magnetic charges. It cannot display electric nor magnetic net
charge in the sphere $S^k(R)$ which links both tubes.
\end{example}

Electric and magnetic fields are well-defined throughout the above
dipole: field lines going into one mouth of the wormhole and coming
out of the other mouth create pairs of equal and opposite charges. One
disadvantage of this construction for modelling particles is that each
charge is connected to one distant opposite charge - apparently in
contradiction to the observed indistinguishability of particles
\cite{sorkin}.

\begin{example}[Magnetic monopole with quadrupole]
\label{magmonoquadru}
Sorkin~\cite{sorkin} used the identity for both identifications:
$\phi(t)=t$ and $\psi(\vec r)=\vec r$. This wormhole is
non-space orientable, but time orientable. It naturally carries a
Faraday two form induced by two equal magnetic charges, leading to a
monopole with additional higher moments.
\end{example}

Note that in order to have a non-zero net magnetic charge it is
necessary for \emph{every} spanning $\Sigma$ with $S^2$ as its
boundary to be non-orientable. A wormhole which appeared and then
disappeared at $t = \tau$ could not display net charge. This can be
seen by applying Stoke's theorem to the orientable three manifold $S^2
\times [0,\tau] \cup B^3_{t=\tau}$. The fact that this three manifold
is timelike in places is irrelevant to the application of Stoke's
theorem.

A wormhole where all spanning surfaces $\Sigma$ of $S^k(R)$ are
non-coorientable is constructed as follows:

\begin{example}[Electric monopole with quadrupole]
\label{elemonoquadru}
We use a reflection for the space identification $\psi({\vec
r}):={\vec r} -2{\ip{{\vec r},a}\over \ip{a,a}}a$ and a reflection for
the time identification $\phi(t)=-t$ (figure~\ref{fig:wormtube}). This
wormhole naturally carries a static Faraday bi vector induced by two
equal electric charges, leading to a monopole with an additional
higher moment. It may also carry a Faraday two form which exhibits a
magnetic dipole moment as in the case of Wheeler's wormhole.
\end{example}

The result that non-time orientable wormholes have net electric
charge but not magnetic charge has added significance because it is
known \cite{hadley97} that the lack of a time orientation could be
used to explain quantum phenomena. By contrast there is no physical
reason in support of (or contrary to) the existence of spacetimes
which do not have a space orientation.

The example constructed possesses closed timelike curves. Although
spacelike slices can be taken they are not hypersurfaces, in the sense
that each point of the surface is crossed once and once only by a
timeline through each point of $M$. The last example of this section
can display both types of net charge:

\begin{example}[Electric and magnetic monopoles with quadrupoles]
\label{elemagmonoquadru}
If one uses the i\-den\-ti\-ty for the space identification
$\psi(\vec r)=\vec r$ and a reflection for the time identification
$\phi(t)=-t$, the spanning surface is neither orientable nor
coorientable.
\end{example}
\section{Line bundle construction}\label{linebundle}
The previous method for constructing $4$-geons with interesting
topologies has a limited repertoire plus the problem of establishing
that the structures are indeed smooth manifolds. A sphere with a point
removed is diffeomorphic to $\R^2$, the stereographic projection is
the diffeomorphism, and the point removed corresponds to
infinity. This technique allows any compact manifold to be transformed
into a non-compact one which includes a region of non-trivial
topology. Time will be modelled as a vector bundle - for one time
dimension this corresponds to attaching a timeline at every point on
the manifold. The line bundle is trivial if and only if the spacetime
is time orientable.

To construct other examples of $n$-geon topologies as defined
in~\ref{topological4geon}, recall that a vector bundle $L$ of rank
$(n-k-1)$ over some $(k+1)$-di\-men\-sional manifold $Q$ is a
$n$-dimensional manifold with a smooth projection $\pi\from L\to Q$,
such that the preimages of points $q\in Q$ are real
$(n-k-1)$-dimensional vector spaces $L_q:=\pi^{-1}(q):=\setof{l\in
L\property \pi(l)=q}$. The manifold $Q$ is immersed into $L$ as zero
section. In the physically relevant case $k=2$ and $n=4$ the total
space $L$, the base space $Q$ and the fibres $L_q$ can be thought of
as spacetime, space and worldlines through $q$ respectively.

\begin{proposition}[Special class of $n$-geon] \label{special4geon}
Let $Q$ be some compact $(k+1)$-di\-men\-sional manifold and let $L\to
Q$ be some real vector bundle of rank $(n-k-1)$ over $Q$. Now choose a
point $q_{\infty}\in Q$ and define $M$ by removing the fibre
$L_{q_{\infty}}$ from $L$. Then $M:=L\setdif L_{q_{\infty}}$ defines a
topological $n$-geon.
\end{proposition}
\proof To see that $M$ satisfies the properties of the $n$-geon
definition~\ref{topological4geon} we remark that the vector bundle $L$
is locally trivialisable: (Notation: for any $U\subset Q$ define
$L\restrto{U} :=\pi^{-1}(U) =\setof{L_q\property q\in U }$). Around
$q_{\infty}$ we can choose a small closed disc $D\subset Q$ ($D$ is
diffeomorphic to a $(k+1)$-ball in $\R^{k+1}$) over which $L$ becomes
trivial, i.e.~$L\restrto{D}\isomorph \R^{n-k-1}\cross D$ as vector
bundles. Then $T:=L\restrto{Q\setdif D}\subset M$ defines a
worldtube. Indeed, $M\setdif T = L\restrto{D\setdif\setof{q_{\infty}}}
\isomorph \R^{n-k-1}\cross(D\setdif\setof{q_{\infty}})$ is
diffeomorphic to $\R^{n-k-1}\cross (\R^{k+1}\setdif B^{k+1})$ as
required. The $k$-sphere $S^{k}:=\partial{D}\subset M$ links $T$ and
is clearly the boundary of $\Sigma:=Q\setdif{\interior{D}}$, which is
a compact $(k+1)$-surface through $T$.  \qed

The above special class of wormhole topologies \ref{special4geon}
reduces for $k=2$ and $n=4$ to a real line bundle $L\to Q$ (space
time) over a compact three manifold $Q$ (space) with one worldline
removed $M:=L\setdif L_{q_\infty}$. This removed worldline plays the
role of the observer at infinity.

As with the topological definition of wormholes, the line bundle
construction has been designed to facilitate the addition of a
Lorentzian metric but it does not carry a metric (see remark~
\ref{rem:nometric}). If a Lorentzian metric is added to the line bundle
then the definition of a space and time orientation should coincide
with the following definition.

\begin{definition}[Space and time orientability]
\label{orientabilitycharacteristics}
If the $n$-geon is given by the above construction $M:=L\setdif
L_{q_{\infty}}$, $M$ is called \emph{space orientable} if $Q$ is
orientable and $M$ is called \emph{time orientable} if the vector
bundle $L\to Q$ is orientable.
\end{definition}

As in the previous section, the vanishing results
\ref{vanishingmagneticcharge} and \ref{vanishingelectriccharge} can be
applied to give:

\begin{proposition}\label{vanishingnetcharge}
If the $n$-geon $M$ is space orientable, then $M$ does not contain
magnetic charge, since the spanning $\Sigma$ is orientable. If $M$ is
time orientable, then $M$ does not contain electric charge, since the
spanning $\Sigma$ is coorientable.
\end{proposition}

\begin{example}[Minkowski space]
Minkowski four space is easily rediscovered by taking the trivial line
bundle over the three sphere: $Q=S^3$ and $L:=\R\cross S^3$ since
stereographic projection from a point $q_\infty\in S^3$ defines a
diffeomorphism $L\setdif L_{q_\infty}\isomorph{\R\cross \R^3}$. It
cannot contain net charge.
\end{example}

\begin{remark}[Wormhole examples]
The examples of the previous sections are rediscovered as line bundle
constructions as follows: Wheeler's wormhole example \ref{dipol} is
based upon the trivial line bundle $L:=\R\cross Q$ over $Q:=S^1\cross
S^2$. Sorkin's wormhole example \ref{magmonoquadru} is a trivial line
bundle $L:=\R\cross Q$ over a three dimensional Klein bottle, i.e.~a
non-orientable twofold subquotient of $S^1\cross S^2\to Q$. In example
\ref{elemonoquadru} $L$ is the pull back of the M\"obius strip
$\hbox{M\"o}\to S^1$ to $Q:=S^1\cross S^2$. In the last example
\ref{elemagmonoquadru} $L$ is the pull back of the M\"obius strip to
the three dimensional Klein bottle.
\end{remark}

The next aim is to use the line bundle construction to produce a
spacetime carrying an electric monopole with no higher moments.
Recall, that a topological $4$-geon $L\setdif L_{q_\infty}$ can
contain electric net charge only if the manifold is not time
orientable, i.e.~only if the line bundle is non-trivial
\ref{vanishingnetcharge}. Since $S^3$ is simply connected, all line
bundles over $S^3$ are trivialisable and the bundle constructions
gives back Minkowski space. The simplest quotient of $S^3$ is
\emph{real projective three space} $Q:=\R P^3=S^3/\setof{\pm\id}$. It
can be viewed as the set of real lines through the origin in four
space: $\R P^3 =\setof{\R x \property x
\in\R^4\setdif\setof{0}}$. Recall, that the \emph{canonical line
bundle} $L\to \R P^3$ has as fibre over $\R x$ the line $L_{\R x}:=\R
x$ itself.

\begin{example}[Electric monopole]
The canonical line bundle $L\to\R P^3$ over real projective three
space is non-trivial and the resultant $4$-geon $M:=L\setdif
L_{q_\infty}$ is a simple spacetime obtained by the bundle
construction that can carry electric net charge.
\end{example}

To see how the static electric Coulomb field of an electric charge
$q_e\in\R$ defines a smooth bi vector density on $M$, we do a variation
of the wormhole construction of the previous section. Similar to the
first example \ref{elechargeincommfd} we deal with Minkowski space
with an open solid worldtube removed: $\R\cross\R^{\ge 1}\cross
S^2$. The radial field $E(t,r,\sigma)=1/(4\pi r^2)(0,1,0)$ together
with the observer field $N(t,r,\sigma):=(1,0,0)$ induce a smooth
Faraday bi vector $H:= q_e E \wedge N$. The electric charge contained
in the two sphere $S^2$ linking the removed worldtube corresponds to
$q_e$. Now we identify events on the cylindrical boundary by flipping
the time coordinate (time identification) and using the antipodal map
(space identification): $(t,1,\sigma)\sim (-t,1,-\sigma)$. The smooth
structure should identify the velocity vector $(\dot t,\dot r,\dot
\sigma)$ at $(t,1,\sigma)$ with $(-\dot t,-\dot r,-\dot \sigma)$ at
$(-t,1,-\sigma)$ (mirror image rule). Notice, that neither the static
observer vector field $N$ nor the radial field $E$ are well defined
with these identifications. However, the observer independent wedge
product $H:=q_e N \wedge E$ clearly is.

The one point compactification of $\R^{\ge 1}\cross S^2$ with
${q_\infty}$ (using stereographic projection) can be identified with
the upper half three sphere. The identification on the equator are
given by the antipodal map, showing that the underlying $Q$ can be
identified with real projective three space. The time line bundle
$L\to Q$ must be the non-trivial canonical bundle.

The $\O(3)$ symmetry of the Coulomb field carries over to the bundle
construction.

This construction is simpler than the wormholes of
section~\ref{wormholetopology} and has higher symmetry. Essentially a
ball is removed from space (a cylinder from spacetime) and then joined
up by identifying opposite points, but swapping the time
direction. Externally this is spherically symmetric and has the same
electric field as a point electric charge. The Faraday tensor is
well-defined throughout. Any path through the reidentified region
undergoes a time reversal and a reversal of the radial electric
field. Unlike the wormhole examples this is a simple electric
monopole, with no higher moments.

A similar construction to create a magnetic monopole is not possible:

\begin{remark}
The Lefschetz fixed point theorem shows that the three sphere $S^3$
can only cover orientable three manifolds $S^3\to Q$. Hence, the line
bundle construction based upon a discrete subquotient $Q$ of $S^3$ can
never carry a magnetic monopole. Similarly, for the magnetic monopole
field from example \ref{magchargeincommfd}, we cannot find an
identification free of fixed points linking the two sphere $S^2$ with
itself, which leaves the magnetic monopole two form invariant. The
Lefschetz fixed point theorem determines such an identification to be
orientation reversing, which rules out the possibility of a magnetic
monopole.
\end{remark}
\section{Conclusion}
New topological structures for spacetime are given which can exhibit
an apparent net electric charge without any apparent
source. Spacetimes with non-orientable immersed surfaces were known to
exhibit magnetic charge (in the sense of definition
\ref{magneticcharge}). The spaces with non-coorientable immersed
surfaces described exhibit the opposite type of charge (electric by
definition \ref{electriccharge}). These spacetimes are not time
orientable; this may seem unphysical, but they are the type of
classical structure which would be required to exhibit quantum
mechanical effects \cite{hadley97}. So the classical, gravitational,
model for quantum mechanics is also seen to lead naturally to the
existence of electric charge and the absence of magnetic charge.
\ack
The first author would like to thank the Deutsche
Forschungsgemeinschaft for financial support through the SFB 256 {\em
Nichtlineare partielle Differentialgleichungen}.
\setlength{\unitlength}{0.8mm}
\begin{figure}[p]
\center{

\begin{picture}(100,200)(40,0)
\linethickness{0.5mm}
\renewcommand{\yscale}{0.4}
\scaleput(60,140){\bigcircle{40}}
\scaleput(140,80){\bigcircle{40}}
\scaleput(100,110){\bigcircle{160}}
\scaleput(20,10){\dashbox(0,100){}}
\scaleput(180,10){\dashbox(0,100){}}
\scaleput(130,190){\makebox(0,0)[bl]{$S^2(R)$}}

\linethickness{0.2mm}
\scaleput(40,10){\dashbox(0,100){}}
\scaleput(80,10){\dashbox(0,100){}}
\scaleput(120,10){\dashbox(0,100){}}
\scaleput(160,10){\dashbox(0,100){}}
\scaleput(40,10){\vector(0,1){10}}
\scaleput(42,15){\makebox(0,0)[bl]{$t$}}

\scaleput(60,140){\circle*{1}}
\scaleput(60,140){\line(1,0){18}}
\scaleput(60,140){\line(2,1){12}}
\scaleput(59,139){\makebox(0,0)[t]{$-a$}}
\scaleput(62,146){\makebox(0,0)[bl]{${\vec r}$}}

\scaleput(75,155){\makebox(0,0)[bl]{$A =(t,{\vec r}-a)$}}
\scaleput(50,120){\makebox(0,0)[tr]{$S^2$}}
\scaleput(72.14,155.14){\circle*{2}}

\scaleput(140,80){\circle*{1}}
\scaleput(140,80){\line(1,0){18}}
\scaleput(140,80){\line(-2,1){12}}
\scaleput(139,79){\makebox(0,0)[t]{$+a$}}

\scaleput(125,95){\makebox(0,0)[br]{$(\phi(t),\psi({\vec r})+ a)=A'$}}
\scaleput(127.86,95.14){\circle*{2}}

\end{picture}
}
\caption{A generalized wormhole construction with one space dimension
suppressed. Unit world tubes are removed from $\R^3$; points on the
boundaries $S^2$ are identified according to the orthogonal maps
$\psi$ and $\phi$. Net charge is defined to be contained in the
outer two sphere $S^2(R)$, linking the large worldtube.}
\label{fig:wormtube}
\end{figure}
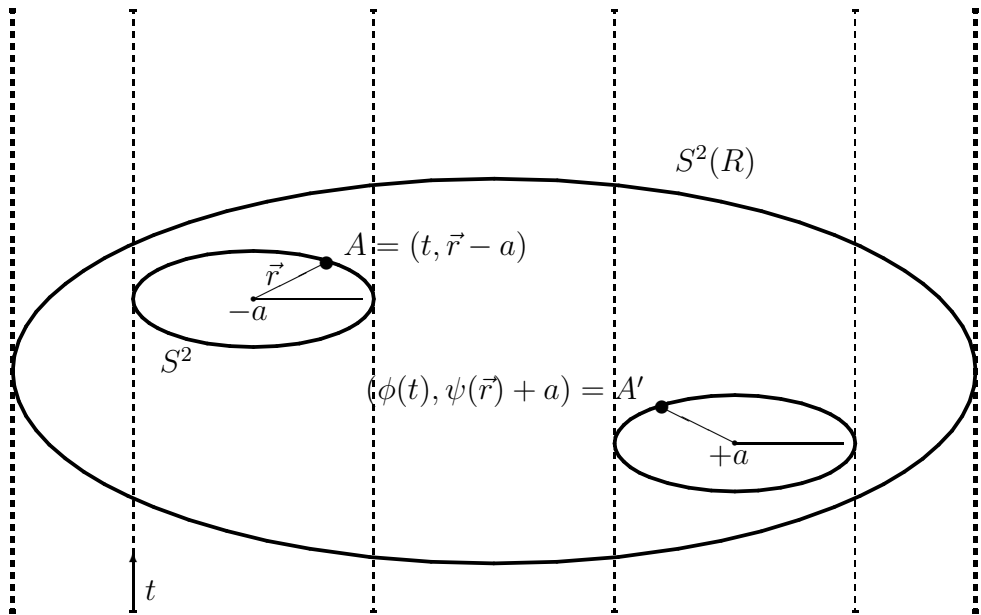

\appendix
\section{Densities and orientations}
In this appendix we recall the basic definitions from the theory of
integration along submanifolds.

\begin{definition}[Density]\label{densities}
Let $V$ be a real $n$-dimensional vector space (here
$V$ will play the role of the tangent space at a point of a manifold
or submanifold). A homogeneous map $\mu \from \Lambda^nV
\setdif\setof{0} \to \R$ with the property $\mu(\lambda\omega) =
|\lambda| \mu(\omega)$ for all $\lambda \in \R\setdif\setof{0} $ and
$\omega\in\Lambda^nV$ is called a \emph{density} (of weight $-n$).
\end{definition}

The set of all densities builds a one dimensional oriented vector
space denoted by $L^{-n}(V)$ or simply $L^{-n}$. A positive density
assigns naturally a real number as volume to an $n$-dimensional
parallelepiped. The tensor product $L^{-n} \tens \Lambda^nV$ is the
space of \emph{pseudo scalars}. This one dimensional space naturally
carries a norm given by $||\mu \tens \omega||:=|\mu(\omega)|$. Using
the induced inner product, the dual space $L^{n} \tens
\Lambda^nV^\dual$ can naturally be identified with $L^{-n} \tens
\Lambda^nV$. The two orientations of $V$ are in one to one
correspondence with the two unit elements of $L^{-n} \tens
\Lambda^nV$.

The above algebraic constructions can be done pointwise on any
$n$-dimensional manifold $M$. In particular at $x\in M$ we simply
write $L^{-n}_x:=L^{-n}(T_xM)$. The resulting line bundle $L^{-n}\to
M$ is trivialisable, but does not come with a canonical
trivialisation. The sections of $L^{-n}$ can naturally be integrated
over $M$.

\begin{proposition}[Integration over oriented submanifolds]
\label{integrationoverorsubmfds}
Let $S\immersion M$ denote a $k$-dimensional immersed submanifold. If
$S$ is orientable and $\orie\in L^{-k}\tens \Lambda^kTS$ a choice of
orientation then any $k$-form $F$ on $M$ can be turned into a density
$\ip{F,\orie}$ over $S$, which can therefore be integrated naturally.
\end{proposition}

On every manifold $M$ we have the deRham sequence of exterior
derivatives on differential forms $d\from C^\infty(M,\Lambda^kT^\dual
M) \to C^\infty(M,\Lambda^{k+1}T^\dual M)$ as a natural differential
operator.

\begin{theorem}[Stoke's theorem] \label{stokesthm}
If $F\in C^\infty(M,\Lambda^kT^\dual M)$ is
any smooth $k$-form on an $n$-manifold $M$, and $\Sigma \immersion M$
an (immersed) compact oriented $(k+1)$-dimensional submanifold with
boundary $\partial\Sigma$ then $$\int_{\partial\Sigma,\orie} F =
\int_{\Sigma,\orie} d F.$$
\end{theorem}
(for a proof see \cite{abraham_etal}).

We recall that a \emph{coorientation} of an immersed $k$-manifold
$S\immersion M$ is an orientation of the normal bundle (quotient
bundle) $TM/TS \to S$ along $S$. Whether the normal bundle is
orientable or not depends upon how $S$ is immersed into $M$.

\begin{proposition}[Integration over cooriented submanifolds]
\label{integrationovercoorsubmfds}
If $S\immersion M$ is coorientable and $\coorie$ is a choice of
coorientation then any $(n-k)$-multi vector density $H$ on $M$ can be
turned into a density $\ip{H,\coorie}$ on $S$, which therefore can
naturally be integrated.
\end{proposition}
\proof At each point of $S$ this density $\ip{H,\coorie}$ is defined
on $k$-vectors $v\in\Lambda^k T\Sigma$ as $\ip{H,\coorie}(v) :=
\ip{\norie^\dual,H} (\norie \wedge v)$, where $\norie \in
\Lambda^{n-k}(TM/TS)$ is a positive normal (according to the
coorientation), and $\norie^\dual \in \Lambda^{n-k}(TM/TS)^\dual$ is
its dual $\ip{\norie^\dual,\norie} = 1$. This definition is
independent of the choice of $\norie$.  \qed

Adjoint to the deRham sequence is the sequence of exterior divergences
of multi vector density fields $div\from C^\infty(M,L^{-n} \tens
\Lambda^{n-k}TM) \to C^\infty(M,L^{-n} \tens \Lambda^{n-k-1}TM)$. If
$\mu$ denotes a non-vanishing density and $X$ a vector field, the
divergence of $\mu \tens X$ can invariantly be defined as a Lie
derivative: $div (\mu \tens X) := \Lie_X \mu$.  Similarly, on a
decomposable bi vector density field we have $div (\mu \tens X \wedge
Y)=(\Lie_X \mu) \tens Y -(\Lie_Y \mu) \tens X + \mu \tens
[X,Y]$. Exterior derivative $d$ and exterior divergence $div$ are
related using a local orientation $\orie\in L^{-n}\tens \Lambda^nTM$
and its dual $\ip{\orie^\dual,\orie}=1$: an orientation turns a $n-k$
multi vector density into a $k$-form and vice versa: $div H =
\ip{{\orie},{d \ip{\orie^\dual,H}}}$.

\begin{theorem}[Divergence theorem]\label{divergencethm}
If $H\in C^\infty(M,L^{-n}\tens
\Lambda^{n-k}TM)$ is any smooth $(n-k)$-multi vector density on an
$n$-manifold $M$, and $\Sigma \immersion M$ an (immersed) compact
cooriented $(k+1)$-dimensional submanifold with boundary
$\partial\Sigma$ then $$\int_{\partial\Sigma,\coorie} H =
\int_{\Sigma,\coorie} div H.$$
\end{theorem}
(for a proof see \cite{abraham_etal}, also \cite{sorkin}).

\end{document}